# Unveiling Optimal SDG Pathways: An Innovative Approach Leveraging Graph Pruning and Intent Graph for Effective Recommendations

Zhihang Yu[1,3], Shu Wang[2,*], Yunqiang Zhu[2], Wen Yuan[2], Xiaoliang Dai[2,4], Zhiqiang Zou[1,3,4]

[1]College of Computer, Nanjing University of Posts and Telecommunications, Nanjing 210023, China.

[2]State Key Laboratory of Resources and Environmental Information System, Institute of Geographic Sciences and Natural Resources Research, Chinese Academy of Sciences, Beijing, China.

[3]Jiangsu Key Laboratory of Big Data Security & Intelligent Processing, Nanjing University of Posts and Telecommunications, Nanjing 210023, China.

[4]University of Chinese Academy of Sciences, Beijing, China.

* Corresponding author:

Shu Wang ( wangshu@igsnrr.ac.cn )

Address: 11A, Datun Road, Chaoyang District, Beijing, 100101, China;

***Author emails:***

Zhihang Yu: 1222045518@njupt.edu.cn ;

Shu Wang: wangshu@igsnrr.ac.cn ;

Yunqiang Zhu: zhuyq@lreis.ac.cn ;

Wen Yuan: yuanw@lreis.ac.cn ;

Xiaoliang Dai: daixiaoliang079x@igsnrr.ac.cn ;

Zhiqiang Zou: zouzq@njupt.edu.cn ;

**Abstract：**

The recommendation of appropriate development pathways, also known as ecological civilization patterns for achieving Sustainable Development Goals (namely, sustainable development patterns), are of utmost importance for promoting ecological, economic, social, and resource sustainability in a specific region. To achieve this, the recommendation process must carefully consider the region's natural, environmental, resource, and economic characteristics. However, current recommendation algorithms in the field of computer science fall short in adequately addressing the spatial heterogeneity related to environment and sparsity of regional historical interaction data, which limits their effectiveness in recommending sustainable development patterns. To overcome these challenges, this paper proposes a method called User Graph after Pruning and Intent Graph (UGPIG). Firstly, we utilize the high-density linking capability of the pruned User Graph to address the issue of spatial heterogeneity neglect in recommendation algorithms. Secondly, we construct an Intent Graph by incorporating the intent network, which captures the preferences for attributes including environmental elements of target regions. This approach effectively alleviates the problem of sparse historical interaction data in the region. Through extensive experiments, we demonstrate that UGPIG outperforms state-of-the-art recommendation algorithms like KGCN, KGAT, and KGIN in sustainable development pattern recommendations, with a maximum improvement of 9.61% in Top-3 recommendation performance.

# 1. Introduction

The sustainable development pattern, also referred to as the ecological civilization pattern in China, represents a strategic approach towards attaining the sustainable development goals (SDGs), which entails meeting the needs of the present generation while safeguarding the ability of future generations to meet their own needs. The primary objective is to enhance the quality of human life while ensuring that it remains within the carrying capacity of the supporting ecosystem. The recommendation of sustainable development patterns involves a customized process that considers the specific conditions of the target region, including natural factors, economic levels, and cultural development (Abisheva et al., 2018; S. Wang et al., 2021). By doing so, it aims to identify and select appropriate sustainable development patterns that are well-suited for the target region. The implementation of a suitable sustainable development pattern can enhance resource utilization, strengthen the protection of cultural resources, and foster the harmonious development of the economy, society, and ecological preservation within the target region. Therefore, the recommendation of suitable sustainable development patterns serves as a pivotal approach towards achieving SDGs (Bennich et al., 2020; Nhamo et al., 2019).

The task of recommending sustainable development patterns shares similarities with recommendation systems in computer science. Both aim to leverage the historical behavior between users (target regions) and items (sustainable development patterns) to establish connections, uncover latent interests, predict preferences, and select the most suitable items for specific users (Mohamed et al., 2019; Tarus & Niu, 2017).

Recommendation systems have demonstrated effectiveness in various domains such as e-commerce (J. Wang et al., 2018), music (Hu et al., 2018), and movies (Han et al., 2021). However, their performance in the geographic domain is limited due to the unique characteristics of geography, including temporal, spatial, and specific land attributes (Wang et al., 2022; Xu et al., 2022; Zeng et al., 2022). Consequently, traditional recommendation models for goods, movies, and music are not directly applicable to recommending sustainable development patterns. Currently, sustainable development pattern recommendations heavily rely on experts and manual processes. However, as the consideration of geographic environmental factors and the long-term nature of ecological civilization recommendations increase, manual recommendations become increasingly challenging. There is a growing need for scientific, accurate, and automated approaches to recommend sustainable development patterns. This has become an important research topic, as the development of reliable and efficient recommendation methods can significantly contribute to achieving SDGs.

Current recommendation systems applicable to sustainable development pattern recommendations can be classified into three categories. The first category is content-based recommendation, which suggests items to users based on the similarity between items calculated from their content. For example, in the case of movies, features such as genre, actors, directors, etc., can be extracted from metadata to calculate item similarity. However, sustainable development patterns are abstract concepts lacking attribute features found in real-world items. As a result, content-based recommendation methods are unsuitable for recommending sustainable development patterns. The second category is collaborative filtering-based recommendation, which can be divided into item-based collaborative filtering and user-based collaborative filtering. These approaches compute similarity between users or items based on their interaction records to make recommendations (Najmani et al., 2022). However, this approach overlooks the rich geographic features such as economy, culture, and natural resources. It solely relies on the interaction between the target region and sustainable development patterns, disregarding the heterogeneity of geographic features. The third category is knowledge graph-based recommendation. A knowledge graph represents a large-scale semantic network consisting of various real-world concepts as nodes and the relationships between them as edges (Zhu et al., 2022). Knowledge graphs have demonstrated significant value in applications like intelligent question answering, text understanding, and recommendation systems (Balloccu et al., 2023; Chen et al., 2021; Dubey et al., 2018). Each node represents an entity, while each edge represents a semantic relationship, forming a vast semantic network that provides rich semantic information. This allows for better exploration of connections between users and items, addressing the data sparsity and cold-start problems faced by collaborative filtering and content-based filtering methods (Liu & Duan, 2021). Due to the sparsity of geographic space, recommending ecological civilization patterns in geographical scenarios faces more severe sparsity issues. Consequently, knowledge graph-based methods offer a promising approach for recommending sustainable development patterns. Knowledge graph-based recommendation systems can be classified into three types: embedding-based methods, path-based methods, and unified methods (Liu & Duan, 2021), as detailed in relevant works in Section 2.

While knowledge graph-based recommendation systems have achieved success in domains like music, movies, and products, they are currently unsuitable for sustainable development pattern recommendation scenarios. This limitation is evident in the following aspects: First, knowledge graph-based recommendation models construct knowledge graphs consisting of an item graph (H. Wang et al., 2018; Zhang et al., 2016), which connects items with their features or attributes, and a user-item graph (H. Wang, M. Zhao, et al., 2019; X. Wang, X. He, et al., 2019), which links items with their attribute or feature links, as well as user-item interactions. However, these graphs often lack user-specific features. For instance, in the KGCL method (Yang et al., 2022) using the Yelp2018 [1] dataset to construct a knowledge graph, only restaurant-centric features, such as Wi-Fi coverage or all-day availability, are considered for the item graph, disregarding user-specific features (Liu & Duan, 2021). In the context of sustainable development pattern recommendation, this means the neglect of regional features and the failure to consider the heterogeneity of geographical environmental factors. Second, in sustainable development pattern recommendation scenarios, user features comprise both continuous and discrete data, with continuous data being more prevalent. If user feature nodes are treated as entities in the knowledge graph, these continuous data entities are represented as single feature values. After inserting most continuous data entities into the knowledge graph, they only connect with the user entity itself and do not establish connections with other entities to aid in inferring user preferences. Consequently, the associativity within the knowledge graph is relatively low. For example, in cases where the GDP feature between two regions is slightly different, a triple <user1, GDP, user2> may not be formed in the knowledge graph. However, in actual scenarios where the GDP between two regions is not significantly different, assuming the existence of a connection <user1, GDP, user2> can be reasonable. Third, in sustainable development pattern recommendation scenarios, the sparsity of geographical data is evident in the fact that the number of users is much larger than the number of items, resulting in sparser user-item interactions and exacerbating the problem of interaction sparsity in recommendation systems. To summarize, the issues of neglecting user features, low associativity within the knowledge graph, and sparse geographical data render existing knowledge graph-based recommendation methods in computer science unsuitable for sustainable development pattern recommendation. It is crucial to integrate regional development features and adapt knowledge graph recommendation methods accordingly to achieve scientifically accurate recommendations for sustainable development patterns.

---

[1] https://www.yelp.com/dataset/documentation/main

In response to the limitations of existing knowledge graph-based methods, we propose a novel approach called the User Graph after Pruning and Intent Graph (UGPIG) approach. The UGPIG method aims to provide precise recommendations for sustainable development patterns by considering the historical interactions and geographical characteristics of the target region. Here is an overview of the UGPIG approach: Firstly, we focus on the target region and construct a Sustainable Development Pattern Knowledge Graph by gathering geographical features specific to that region. This knowledge graph is referred to as the User Graph (UG). To enhance the UG's associativity, we prune and reconstruct the continuous data entity nodes, resulting in the User Graph after Prune (UGP). This process significantly improves the connectivity between various regional features, enabling better exploration of geographical heterogeneity features. The denser interconnections facilitate the identification of rich features and case studies, thereby supporting more precise recommendations. Secondly, we leverage the user intent method (X. Wang et al., 2021) to create an Intent Graph (IG), which analyzes the historical interaction features and intent features of the target region. The IG provides valuable insights into user preferences and intent related to sustainable development patterns. Finally, we simultaneously explore the geographical features of the target region on the UGP and integrate the information from both the UGP and IG using attention mechanisms. By combining the UGP and IG, the UGPIG method effectively captures the intricate relationships between users, geographical features, and sustainable development patterns. This comprehensive approach overcomes the limitations of existing knowledge graph-based recommendation approaches, allowing for more accurate recommendations that incorporate both user interactions and geographical characteristics. In conclusion, the UGPIG approach offers a novel and effective solution for recommending sustainable development patterns. It addresses the shortcomings of previous knowledge graph-based methods by considering user-specific features, enhancing the associativity within the knowledge graph, and incorporating geographical data. The integration of the UGP and IG allows for a more holistic understanding of user preferences and regional characteristics, leading to scientifically accurate and precise recommendations for sustainable development patterns.

The primary aim of this research is to examine the incubation conditions for sustainable development patterns, provide recommendations for appropriate sustainable development patterns in the target region, and ultimately facilitate regional sustainable development. This study makes two significant contributions:

- We conduct a thorough validation of the impact of User Graph pruning and Intent Graph on the sparsity of geographical data and spatial heterogeneity. Specifically, we design and implement pruning steps for the knowledge graph in the field of geography, ensuring a more refined and representative User Graph. Additionally, we develop the Intent Graph, which effectively captures historical interaction features and user intent, enhancing the sustainable development patterns recommendation performance.
- We propose a novel recommendation method called UGPIG, specifically tailored for the field of environmental geography. This method successfully addresses the sparsity issue among recommended entities and incorporates spatial heterogeneity into the recommendation process. To demonstrate its effectiveness, we apply UGPIG in a case study at the provincial level, showcasing its practical application and establishing a solid theoretical and methodological foundation for future research on intelligent sustainable development pathways.

This paper is structured as follows: Section 2 provides a detailed explanation of the three types of existing knowledge graph-based recommendation methods. Section 3 introduces the main framework and implementation details of our proposed method, UGPIG. In Section 4, we present comprehensive performance experiments conducted to study the impact of model parameters. Section 5 focuses on analyzing the connotation of the intent vector in the Intent Graph and presents a case study conducted in Fujian Province. Finally, in Section 6, we conclude our work, summarizing the key findings and contributions of our research.

**2. Related work**

In this section, we will briefly introduce three recommendation methods based on knowledge graphs. In this paper, we focus on and draw inspiration from the third method, which is the unified methods.

Embedding-based Methods: These methods utilize knowledge graph embedding techniques to map information from the knowledge graph into a lower-dimensional space, enriching the representation of items. There are two types of knowledge graph embedding techniques: translation-based models like TransR (Lin et al., 2015) and TransE (Bordes et al., 2013), and semantic matching models like DistMult (Yang et al., 2014). After learning the embedding information, it is used as prior knowledge to guide the recommendation process and predict user preferences for items. Based on the order of training the knowledge graph feature learning module and the recommendation module, embedding-based methods can be further divided into sequential training, alternating training, and joint training. For example, in the news recommendation model DKN (H. Wang et al., 2018), semantic representations of news articles are integrated through the knowledge graph feature learning module, followed by training the recommendation module. This belongs to the sequential training approach. Similarly, joint training approaches include CKE (Zhang et al., 2016), while alternating training approaches include MKR (H. Wang, F. Zhang, M. Zhao, et al., 2019). Although embedding-based methods are straightforward, they often lack interpretability. The embedding methods map entities and relationships into a low-dimensional vector space, which can be challenging to interpret. In the context of sustainable development pattern applications, interpretability is crucial as it represents the importance of geographical factors.

Path-based Methods: The main idea behind path-based methods is to construct a user-item graph and leverage the connectivity between nodes in the graph to enhance recommendation performance. Depending on the usage of paths, path-based methods can be divided into two categories. The first category utilizes semantic similarity between users and users, users and items, and items and items under different meta-paths to regularize the representation of users and items and calculate user preferences for items. Representative algorithms in this category include RuleRec (Ma et al., 2019). The second category focuses on explicitly embedding user-item path information and directly models the user-item relationship. For a user $u$ and an item $i$, assuming there are $n$ paths, each path can be represented as a vector through learning. Finally, by integrating the vector representations of all paths, user preferences or ratings for items are computed based on the representations of users, items, and paths. Typical algorithms in this category include KPRN (X. Wang, D. Wang, et al., 2019). Compared to embedding-based methods that only utilize first-order connectivity in the knowledge graph, path-based methods have the advantage of mining higher-order connectivity information in the graph between users and items. However, path-based methods also have some drawbacks. For example, the design of meta-paths in terms of size and types relies on manual and extensive domain knowledge, which can be time-consuming. In the context of sustainable development pattern recommendation in geoscientific scenarios, the manual design of meta-paths is costly and not suitable for long-term usage.

Unified Methods: unified methods have been a recent research trend, and our UGPIG method, along with three baseline models (KGCN (H. Wang, M. Zhao, et al., 2019), KGAT (X. Wang, X. He, et al., 2019), and KGIN (X. Wang et al., 2021)), falls into this category. To utilize the semantic information from embedding-based methods and the connectivity information from path-based methods, unified methods combine them to fully leverage the information from both sides for better recommendation performance. For example, KGCN aggregates the domain structure of entities in the knowledge graph and utilizes multi-hop neighbor structures of items to enrich the representation of entities and transmit high-order node information. Moreover, the process of propagating high-order information in joint-based methods can be seen as a process of discovering user preferences (Liu & Duan, 2021), providing the model with a certain level of interpretability.

## 3. Method

### 3.1 Problem Formulation

In the recommended scenario of sustainable development patterns, let $U$ be a set of target regions, and $I$ be a set of sustainable development patterns. The interaction matrix $Y \in \mathbb{R}^{|U| \times |I|}$ represents the historical interaction between target regions and sustainable development patterns. If $y_{u,i} = 1$, it indicates that target region $u \in U$ has previously adopted sustainable development pattern $i \in I$, whereas $y_{u,i} = 0$ indicates no historical interaction between region $u$ and sustainable development pattern $i$. Additionally, UGP refers to the knowledge graph of the target regions, which includes a set of entities $\mathcal{V}$ and a set of relationships $\mathcal{R}$ (for specific examples, refer to Section 3.2.1). Through the IG, which connects target regions with their past adopted sustainable development patterns, the intent common intent vector set $P$ is used. Here, $p$ represents an individual intent vector.

Our goal is to learn a function $\mathcal{F}$, given the provided UGP, IG, and interaction matrix $Y$, to predict the score $\hat{y}_{u,i}$ for a target region $u$ and a sustainable development pattern $i$. By comparing the scores, we can obtain the top-K sustainable development patterns for the target region:

$$\hat{y}_{u,i} = \mathcal{F}(u, i | UGP, IG, Y) \tag{1}$$

### 3.2 Framework overview

Our proposed UGPIG method, as illustrated in Figure 1, consists of three layers: the Graph Construction Layer, the Information Aggregation Layer, and the Feature Fusion Layer. First, in Graph Construction Layer, considering the heterogeneity of geographical features among different regions, we employ techniques such as natural language processing and web text mining to gather sustainable development pattern case studies (representing historical interactions between target regions and sustainable development patterns) and geographical features of the target regions from various data sources such as official websites and news media. We construct the knowledge graph UG using the geographical features of the target regions and build the IG using the case studies of sustainable development patterns. We then prune and reconstruct UG to generate UGP, enhancing the relevance of the knowledge graph (details of the pruning operation can be found in Section 3.3.1). Next, in the Information Aggregation Layer, we aggregate information on UGP and IG separately. This involves extracting heterogeneous features of the target regions, historical interactions, and intent features. Finally, in the Feature Fusion Layer, we employ attention mechanisms to fuse the two sets of features obtained from the information aggregation layer. This approach considers both types of features and produces a representation of the target region that addresses data sparsity. By taking the dot product between the target region representation and the representation of sustainable development patterns obtained from the aggregation module, we obtain preference scores for the target region regarding the various sustainable development patterns.

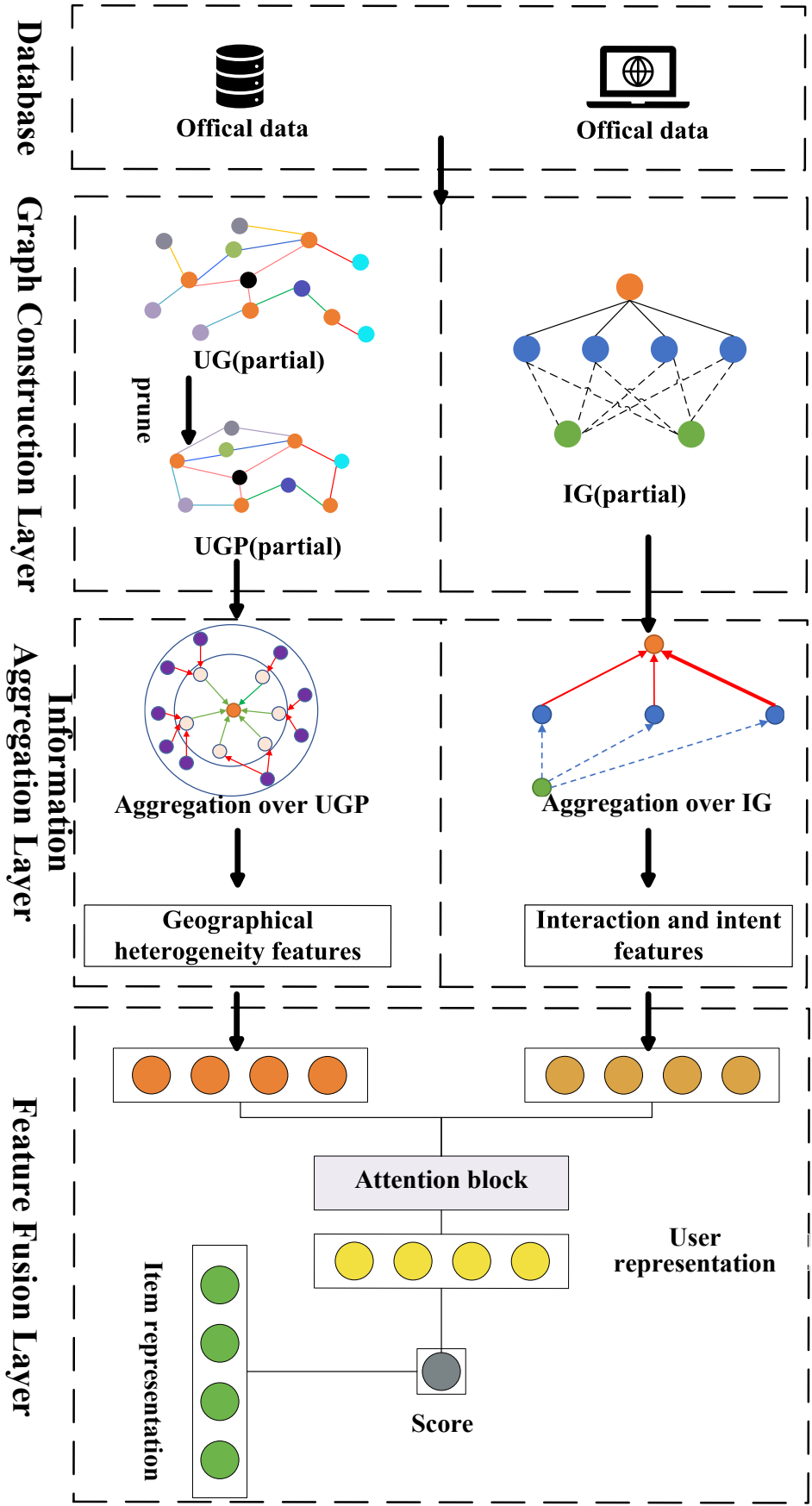

Figure 1 The structure of the UGPIG.

## 3.3 Construction of Graphs

To recommend sustainable development patterns, it is essential to construct a comprehensive semantic network that encompasses various resource information of regions and prominent sustainable development patterns, and establish their associations. Therefore, in this part, we first construct UG based on the information available in the database. Then, we prune and reconstruct the continuous features from the user's knowledge graph to form UGP. Subsequently, we utilize historical interaction data and manually defined intent vector group $P$ to build IG.

### 3.3.1 Construction and pruning of User Graph

As shown in Figure 1, the Database represents the two approaches through which we collect the necessary data: the web and official sources. Such as Baidu Encyclopedia and Resource and Environment Science and Data Center of Chinese Academy of Sciences[2]. We gather a vast amount of sustainable development pattern data from these two sources. After aggregating, aligning, and disambiguating the collected data, we obtain structured data required to build the UG and IG. In this process, UG is constructed using the geographical feature data of the target regions. UG is a semantic network composed of triples <target region, geographical feature, target region>. It does not include any sustainable development pattern information; it is solely composed of the target regions and their corresponding features.

**User Graph:** The User Graph is centered around the target regions, taking into account the heterogeneity of geographical features across different regions. To address this requirement, six categories of indicators have been scientifically designed to represent the geographical features of the target regions. Specifically, these categories include 8 indicators related to the geographical location, 4 indicators related to the richness of resources, 5 indicators related to economic development, 5 indicators related to environmental friendliness, 3 indicators related to social development, and 4 indicators related to cultural factors. In total, there are 29 indicators, as detailed in Table 1. For a specific target region, various geographical features such as latitude and longitude, altitude, meteorological data, hydrological information, and soil characteristics are collected to construct the User Graph. Each regional entity is connected to these features through the relationships depicted in Table 1. Paths in the form of <target region, geographical feature, target region> are formed among different regions. These paths constitute a large semantic network. By mining the semantic information in the User Graph, we can fully consider the geographical features of the target regions and provide intelligent analysis and decision support for sustainable development pattern recommendations to individuals.

---

[2] https://www.resdc.cn/

Table 1 Data information in User Graph.

| Category | Feature | Type Of Feature | Relation in User KG |
|---|---|---|---|
| Geographical Location | Province | discrete | Area.Province |
| | City | discrete | Area.City |
| | Division | discrete | Area.Division |
| | Precipitation | continuous | Area.Precipitation |
| | Soil | discrete | Area.Soil |
| | Climate | discrete | Area.Climate |
| | Altitude | continuous | Area.Altitude |
| | Landform | discrete | Area.Landform |
| Richness of Resources | Cultivated Area | continuous | Area.CultivateArea |
| | Grass Coverage | continuous | Area.GrassCoverage |
| | Per Capita Water Resources | continuous | Area.WaterPer |
| | Forest Coverage | continuous | Area.WoodCover |
| Economic Development | Gross GDP | continuous | Area.GDP |
| | GDP Per Capita | continuous | Area.GDPPer |
| | Per Capita Savings of Urban Residents | continuous | Area.UrbanSaving |
| | Proportion of Secondary Industry Output Value | continuous | Area.SecondInGDP |
| | Proportion of Tertiary Industry Output Value | continuous | Area.ThirdInGDP |
| Environmental Friendliness | Environmental Quality of Surface Water | continuous | Area.WaterQuality |
| | Soil Erosion Modulus | continuous | Area.SoilErosion |
| | Number of Animal and Plant Habitats | continuous | Area.BiologyHabitat |
| | Proportion of Nature Reserve Area | continuous | Area.NatureReserve |
| | Comprehensive Risk of Natural Disasters | continuous | Area.NatureRisk |
| Social Development | Highway and Railway Density | continuous | Area.HighDensity |
| | Number of Beds in Medical and Health Institutions | continuous | Area.BedsInMedical |
| | Number of Tourist Attractions above AAA | continuous | Area.AAA |
| Cultural Factors | Nationality Group | discrete | Area.Nationality |
| | Dialects | discrete | Area.Dialect |
| | Number of Intangible Cultural Heritage | continuous | Area.NOICH |
| | Intangible Cultural Heritage Type List | discrete | Area.ICHTL |

**User Graph After Pruning:** Although we have obtained a knowledge graph for sustainable development patterns, it is evident that the correlation degree of nodes in UG is very low. This is due to the fact that, as shown in Table 1, 68.97% (20/29) of the user features consist of continuous data, and these continuous features in the knowledge graph often only connect with the target region itself. To improve the correlation degree of the knowledge graph, we can prune the continuous entity nodes from the knowledge graph and reintroduce them through a discretization process to reconstruct the knowledge graph and enhance its correlation degree. This leads to the generation of a new knowledge graph called UGP. As shown in Figure 2, the orange nodes represent target region nodes, while the nodes of other colors represent feature nodes of the target regions. Edges of the same color indicate the same relationship type in UG, and nodes of the same color represent nodes of the same type in UG. Taking the nodes in the figure as an example, some nodes belong to the same category of continuous features. Through the discretization process, it is discovered that these two nodes belong to the same level. Consequently, these two nodes can be pruned from the graph, and new nodes are created to connect with the target regions previously linked to the pruned nodes. The graph density before pruning is 14.29%, while after pruning, the density increases to 27.28%. Therefore, the correlation degree of the knowledge graph is enhanced as originally disconnected target regions now have new paths connecting them. Refer to Algorithm 1 for the specific pruning algorithm.

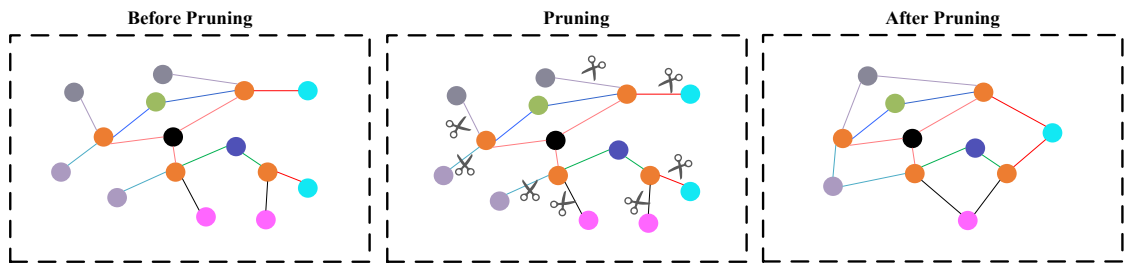

Figure 2 Part of the pruning process of UG.

| | |
|---|---|
| Algorithm 1 User Graph Pruning | |
| Input: | UG: User Graph |
| Output: | UGP: User Graph after Pruning |
| 1. | for $(u,v)$ in UG: |
| 2. |   if $v$ is continuous: |
| 3. |     prune($v$) |
| 4. |     if Discrete($v$) exist: |
| 5. |       connect($u$, Discrete($v$)) |
| 6. |     else: |
| 7. |       v=CreateDiscreteNode($v$) |
| 8. |       connect($u$, $v$) |
| 9. |   else: |
| 10. |     continue; |
| 11. | end for |

For each pair of target region node $u$ and its feature node $v$ in the knowledge graph, Algorithm 1 first checks if the feature $v$ of $u$ is continuous. If it is continuous, the algorithm prunes the feature node $v$ from the graph. Then, it checks if the discretized node Discrete($v$) corresponding to the continuous node $v$ exists. If it exists, the algorithm directly connects the target region $u$ to the new discretized node Discrete($v$). If it does not exist, the algorithm uses the CreateDiscreteNode function to create a new discrete node $v$ and connects node $u$ to node $v$. Finally, if the node $v$ itself is already a discrete feature, the algorithm moves on to the next pair $(u, v)$ of target region and feature. The CreateDiscreteNode function aims to discretize continuous data by dividing it into different levels. Taking the example of the number of animal and plant habitats (ranging from 0 to 35), it can be divided into four levels: level 1 (0), level 2 (1-2), level 3 (3-4), and level 4 (above 5), based on the quantity of habitats. Similar approaches can be applied to discretize other continuous data as well.

### 3.3.2 Construction of Intent Graph

**Intent Graph**：As shown in the Graph Construction Layer in Figure 1, the IG is composed of the sustainable development patterns observed in the target regions and a shared collection of intent vectors among all target regions. In this context, the target region node $u$ (represented as an orange node) is connected to the shared intent vector node $p$ (represented as a blue node) and the sustainable development pattern node $i$ (represented as a green node), forming the relationship <$u, p, i$>. The purpose of IG is to represent the dependency of relationships within the UGP using intent vectors $p$, analyze the connections between the target regions and sustainable development patterns, and provide a certain level of interpretability for the model. It is important to note that the collection of intent vectors $P$ consists of more than one vector, as they are shared among all target regions. Unlike the approach described in KTUP (Cao et al., 2019), where each intent vector is bound to a relationship in the knowledge graph, each intent vector $p$ in the IG is defined as a combination of all relationships in the UGP. This design allows the intent vectors to comprehensively represent the features and needs of the target regions, enabling a better capture of the associations between the target regions and sustainable development patterns.

The intent vector $p$ is defined as follows:

$$e_p = \sum_{r \in \mathcal{R}} \alpha(r,p) e_r \tag{2}$$

The embedding representation of intent vector $p$ is denoted as $e_p$, $\mathcal{R}$ represents the set of all relationships in UG, $\alpha(r,p)$ represents the importance of relationship $r$ to intent vector $p$, $e_r$ represents the embedding representation of relationship $r$ in the UGP, and $w_{rp}$ is a trainable matrix representing the weight between $r$ and $p$. The calculation of $\alpha(r,p)$ is as follows:

$$\alpha(r,p) = \frac{\exp(w_{rp})}{\sum_{r' \in R} \exp(w_{r'p})} \quad (3)$$

Furthermore, each intent vector $p$ is independent of each other, indicating that each intent has different weights on the relationships in UGP. Each intent vector $p$ represents a distinct intent, and there is a shared set of intent vectors among all target regions. Moreover, the degree of reliance on a specific intent vector p may vary across different regions. As shown in Figure 3, assuming there are three vectors in the shared set of intent vectors, the magnitude of $\beta$ determines which intent is more important to the user. Each intent vector $p$ has different weights on the relationships $r$ in UGP. For more information about IG, please refer to (X. Wang et al., 2021).

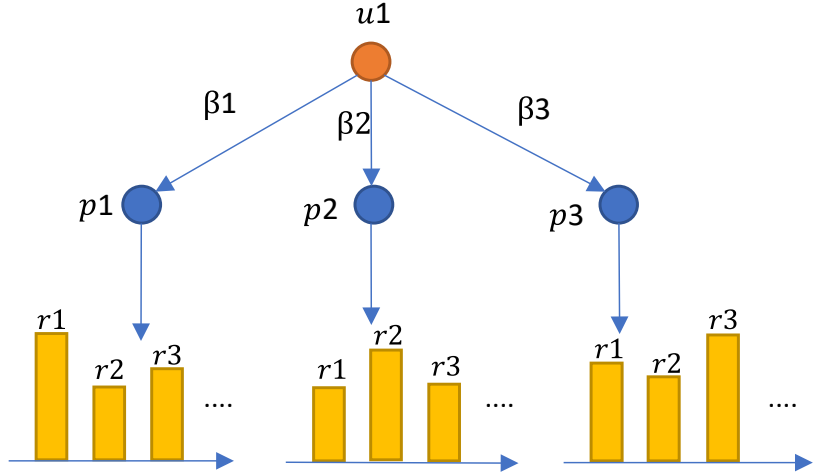

Figure 3 Interpretation of the intention vector.

### 3.4 User Information Aggregation

In this section, based on the semantic networks UGP and IG constructed in Section 3.4, we first aggregate information on UGP to obtain the heterogeneous features of the target region. Then, we aggregate information on IG to capture the historical interactions and intent features of the target region. Finally, these two types of feature vectors, representing different semantic spaces, are fed into the next layer, the Feature Fusion Layer, to obtain the final representation of the target region.

**3.4.1 Aggregation over UGP**

First, we aggregate the geographic features of the target region on UGP to enhance its representation. For a triple $<u, r, v>$ in the UGP, we denote $N_u^{kg}$ as the set of relation-entity pairs $(r, v)$ that are directly connected to $u$ in UGP, where $N_u^{kg} = \{(r,v)|(u,r,v) \in UGP\}$. This represents the first-order connections between the target region and its feature entities. Next, we integrate the information of the relations $r$ and entities $v$ that are first-order connected to u to enrich the representation of $u$:

$$e_u^{kg(1)} = \frac{1}{|N_u^{kg}|} \sum_{(r,v) \in N_u^{kg}} e_r \odot e_v \quad (4)$$

$e_u^{kg}$ represents the embedded representation of the target region $u$ after single-hop aggregation on UGP. $N_u^{kg}$ is the set of relation-entity pairs $(r, v)$ that are connected to the target region $u$ in UGP. $e_r$ represents the embedded representation of relation $r$ in UGP, while $e_v$ represents the embedded representation of entity $v$ in UGP. The symbol $\odot$ denotes element-wise multiplication between two vectors.

To capture higher-order semantic information, the single-hop aggregation can be extended to include multi-hop aggregation. For the purpose of illustration, we consider the 2-hop scenario shown in Figure 4. In this case, the embedded representation of the target region node $e_u$ can be expressed as follows:

$$e_u^{kg(2)} = \sum_{s \in S^{(2)}} \frac{e_{r1}}{|N_{s1}^{kg}|} \odot \frac{e_{r2}}{|N_{s2}^{kg}|} \odot e_{v2} \quad (5)$$

In the equation, $e_u^{kg(2)}$ represents the 2-hop aggregated representation of the target region node $e_u$. $S^{(2)}$ denotes the set of 2-hop paths from the target region node $e_u$. By performing two layers of information aggregation, the path information $e_{r1}$, $e_{r1}$ from the 2-hop paths, as well as the node information $e_{v2}$ from the intermediate nodes, are all captured by the central node $e_u$. This aggregation method preserves the complete path information in the UGP and enables the exploration of the heterogeneity of geographic features within the UGP. For the aggregation in the L-hop layer of multi-hop aggregation, the final representation can be computed as follows:

$$e_u^{kg} = \sum_{l \in L} e_u^{kg(l)} \tag{6}$$

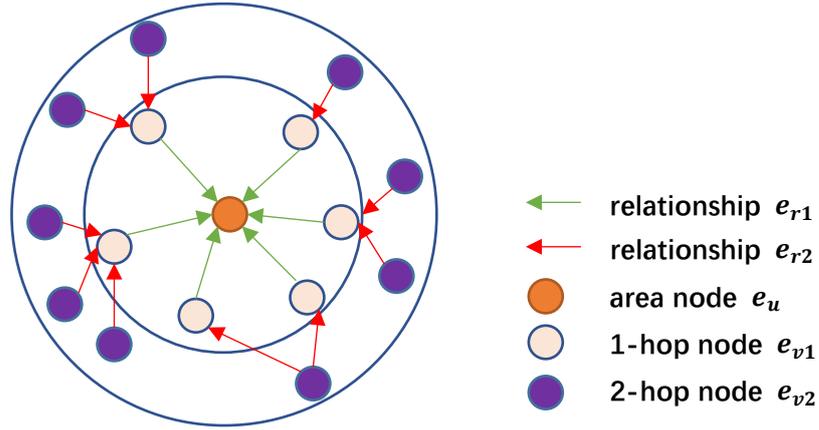

Figure 4 2-hop aggregation of target regions over UGP.

### 3.4.2 Aggregation over IG

In the IG, we aggregate the sustainable development pattern $e_i$ that the target region $e_u$ has experienced in the past to enhance the representation of the user $e_u$ using historical interaction information. Additionally, we consider the intent vector $p$ of the target region as one of the user's features, which also needs to be aggregated onto $e_u$.

The representation $e_u^{IG}$ obtained by aggregating each target region $e_u$ in the intent graph can be expressed as follows:

$$e_u^{IG} = \frac{1}{|N_u^{IG}|} \sum_{(p,i) \in N_u^{Ig}} \beta(u,p) e_p \odot e_i \qquad (7)$$

The representation $e_u^{Ig}$ denotes the embedded representation of the target region $e_u$ after performing single-hop aggregation in the intent graph. $N_u^{IG}$ represents the collection of sustainable development patterns $i$ and intent vectors $p$ associated with the target region $e_u$ in the past. $e_p$ represents the embedded representation of the intent vector $p$. By passing the pattern labels through an embedding layer, we can obtain the embedded representation $e_i$ of the sustainable development pattern $i$. The symbol $\odot$ represents element-wise multiplication between two vectors. $\beta(u,p)$ describes the importance of each intent vector $p$ for the target region $u$.

$\beta(u,p)$ is defined as follow:

$$\beta(u,p) = \frac{\exp(e_p^T e^u)}{\sum_{p' \in P} \exp(e_{p'}^T e^u)} \qquad (8)$$

As shown in Figure 5, the blue dashed line represents the preservation of information from the sustainable development pattern $i$ to the intent vector $p$ through element-wise multiplication. The red solid lines indicate the aggregation of information from the intent vector set to the target region $u$ (thicker lines represent higher importance, and vice versa). In this context, the target region $e_{u1}$ is connected to the sustainable development pattern $e_{i2}$. After the aggregation operation, the representation of the target region $e_{u1}$ incorporates both its historical interactions with sustainable development patterns and all the vectors in the intent vector set $P$ (assuming three intent vectors are defined). This establishes the latent correlations between them, enhancing the representation of the target region $e_{u1}$.

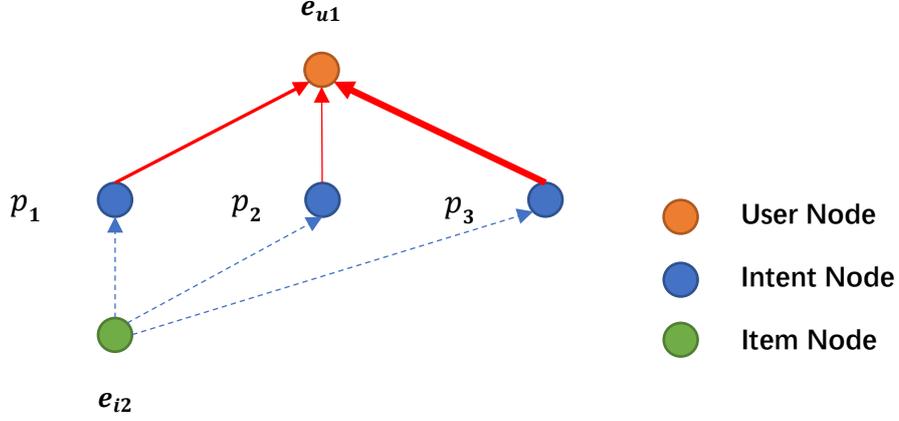

Figure 5 Aggregation of target region u1 on IG.

**3.5 Fusion of features**

In previous section, we obtained the UGP-based feature $e_u^{kg}$ for the target region and the IG-based feature $e_u^{IG}$. Since these two vectors are not in the same semantic space, directly superimposing them cannot yield the final representation $e_u$ of the target region. Therefore, in this section, we employ an attention mechanism to integrate the features from UGP and IG, resulting in the final representation of the target region. Additionally, we calculate the target region's score for each sustainable development pattern.

The target region can eventually be expressed as follows:

$$e_u = \sum_{t \in (KG, IG)} \gamma(t) e_u^t \qquad (9)$$

Whereas, $e_u^t$ represents the feature obtained in the Information Aggregation Layer, and $\gamma(t)$ represents the importance of $e_u^t$. The importance $\gamma(t)$ can be

calculated as follows:

$$\gamma(t) = ReLU\left(\frac{\exp(We^t)}{\sum_{t' \in (KG,IG)} \exp(We^{t'})}\right) \qquad (10)$$

The weight matrix $W$ is a trainable parameter of the model that is used to map the $e_u^{KG}$ and $e_u^{IG}$ onto the same semantic space. It enables the mapping of the two semantic networks onto a shared space. $ReLU$ is a commonly used activation function. The final representation $e_u$ can be obtained by summing the weighted representations from UGP and IG.

After obtaining the representation of the target region $e_u$ and the embedding of the sustainable development pattern $e_i$ through the embedding layer, the objective function can be defined as follows:

$$y'_{u,i} = e_u^T e_i \qquad (11)$$

We calculate the score of the target region $u$ for the sustainable development pattern $i$ by taking the inner product between the representations $e_u$ and $e_i$. This score represents the compatibility or suitability of the target region for the given sustainable development pattern. By comparing the scores of different sustainable development patterns for the target region, we can identify the most suitable pattern for the region's development.

## 4. Experiments

To evaluate the performance of our proposed method UGPIG, we conducted experiments and analyzed the impact of model parameters. Firstly, we present the scale of the experimental data and the evaluation metrics used. Then, we assess the overall performance of the model. Finally, we conduct multiple groups of experiments to investigate the effects of aggregation layers, the number of intent vectors, graph pruning, and attention mechanism on the recommendation performance.

**4.1 Dataset**

Our dataset consists of 94 third-level sustainable development patterns and 2596 target regions. In section 3.3.1, we created User Graph and Intent Graph. The scale of the User Graph is shown in Table 7. Table 2 presents the scale of the Intent Graph, where Interactions represents the number of interactions between target regions and sustainable development patterns.

Table 2 Statistics of the Intent Graph.

|  | Users | Items | Interactions |
| --- | --- | --- | --- |
| Intent Graph | 2596 | 94 | 7830 |

For the dataset, we use 80% of the interactions from target regions as the training set and 20% as the test set. Within the training set, we randomly select 20% as the validation set to adjust the selection of hyperparameters. It is important to note that due to the sparsity of geographical data, some target regions have very few historical interactions and cannot be proportionally divided. For example, if a target region has only two historical interactions, one interaction is allocated to the training set and the other to the test set. If there is only one interaction, it is assigned to the training set to ensure an adequate amount of training data. During training, for each region $u$, the items in the interaction matrix $Y$ where $y_{u,i} = 1$ are considered positive samples. Random negative samples with $y_{u,i} = 0$ are matched to each positive sample for training.

## 4.2 Evaluation criterion

**Evaluation criterion**: To evaluate the Top-K sustainable development pattern recommendations for target regions, we employ the all-ranking strategy (Krichene et al., 2020) instead of the user subset extraction method used in (H. Wang, F. Zhang, M. Zhang, et al., 2019; H. Wang, M. Zhao, et al., 2019). Specifically, for each target region, we recommend sustainable development patterns that have not been interacted with in the past based on the model and select the Top-K patterns that are most suitable for the region's development. We use *Precision@K*, *Recall@K*, and *F1@K* as evaluation metrics to compare the performance of our method with other baseline methods：

$$Precision@K = \frac{|R^K(u) \cap T(u)|}{K} \tag{12}$$

$Precision@K$ measures the proportion of common sustainable development patterns between the Top-K recommended patterns for target region $u$ and the actual interacted patterns $T(u)$ for $u$, among the K patterns.

$$Recall@K = \frac{|R^K(u) \cap T(u)|}{|T(u)|} \tag{13}$$

$Recall@K$ measures the proportion of Top-K recommended sustainable development patterns for target region $u$ that are also present in the true interacted sustainable development patterns $T(u)$ of $u$ in the test set.

$$F1@K = \frac{2 \times Precision@K \times Recall@K}{Precision@K + Recall@K} \tag{14}$$

$F1@K$ is a metric that evaluates the overall performance of a recommendation system by computing the harmonic mean of $Precision@K$ and $Recall@K$.

**Parameter Settings**: For our UGPIG method, we set the embedding dimension of vectors to 64, the batch size to 128, the number of intent vectors to 4, the number of aggregation layers in the User Graph to 4, and the learning rate to 1e-3.

**Baselines**：To validate the effectiveness and feasibility of UGPIG in the context of geospatial sustainable development patterns, we compared its performance with three knowledge graph-based models: KGCN, KGAT, and KGIN.

•**KGCN** extends graph convolutional neural networks to the domain of knowledge graphs. It captures local neighborhood structures by performing aggregation operations on fixed-sized neighborhoods of nodes in the knowledge graph. This enriches the representation of entities in the graph.

•**KGAT** utilizes the knowledge graph embedding technique TransR to obtain initial representations of entities in the knowledge graph. It introduces an attention mechanism in the node aggregation process in the graph, allowing differentiation of the importance of neighboring nodes. This enriches the representations of entities in the knowledge graph based on their varying importance.

•**KGIN** introduces intent vectors to represent different user preferences and proposes a novel path-aware knowledge graph aggregation method to represent entities in the knowledge graph. By incorporating intent vectors and considering the paths in the knowledge graph, KGIN captures the preferences and interests of users and leverages the structural information in the graph to enhance the representation of entities.

## 4.3 Results

### 4.3.1 Comprehensive result

In Section 4.3, we evaluated the performance of the mentioned methods with K=3

and K=5. The experimental results are presented in Tables 3 and 4. It can be observed that the UGPIG method outperforms others in terms of $Precision@K$, $Recall@K$, and $F1@K$ for both Top-3 and Top-5 recommendations.

Taking $F1@K$ and $Recall@K$ as examples, $F1@K$ serves as an overall performance metric for the recommendation system, while $Recall@K$ measures the ability of the model to derive new sustainable development patterns based on the historical interactions between target regions and sustainable development modes. On average, each target region in the IG has three historical interactions. Therefore, we first set K=3 to evaluate the performance of the methods when the number of recommended results is small. In the Top-3 recommendations, $F1@3$ shows an improvement of 9.61% compared to KGCN, 7.02% compared to KGAT, and 3.47% compared to the best baseline model KGIN. $Recall@3$ exhibits the highest improvement of 15.14%, surpassing the best baseline model KGIN by 8.86%. To further validate the effectiveness of our method with a larger number of results, we set K=5. In the Top-5 recommendations, $F1@5$ shows an improvement of 8.53% compared to KGCN, 7.23% compared to KGAT, and 2.16% compared to the best baseline model KGIN. $Recall@5$ exhibits the highest improvement of 19.06%, surpassing the best baseline model KGIN by 7.14%. The experimental results demonstrate that our method performs better regardless of the number of results.

Due to the absence of attention mechanism in the convolutional operation of KGCN, the representation of entities in KGCN incorporates a significant amount of noise. KGAT reduces the interference of noise on entity representation by utilizing knowledge graph embedding technique TransR and attention mechanism, resulting in slightly better performance than KGCN. KGIN enhances entity representation by introducing intent vectors, thereby improving the performance of the recommendation system. Our method, UGPIG, builds upon KGIN by incorporating the exploration of geographic heterogeneity through UGP and enhancing the representation of target regions through graph pruning and attention blocks. This approach achieves the best performance. In summary, the breeding of sustainable development patterns relies primarily on the geographic features of target regions. Baseline models perform poorly because they do not consider the heterogeneity among different regions.

Table 3 Top-3 recommendation result (Bold font is our method).

|       | $Precision@3$ | $Recall@3$ | $F1@3$ |
|-------|---------------|------------|--------|
| KGCN  | 0.0314(-6.93%) | 0.1004(-15.14%) | 0.0478(-9.61%) |
| KGAT  | 0.0537(-4.7%)  | 0.1173(-13.45%) | 0.0737(-7.02%) |
| KGIN  | 0.0820(-1.87%) | 0.1632(-8.86%)  | 0.1092(-3.47%) |
| **UGPIG** | **0.1007** | **0.2518** | **0.1439** |

Table 4 Top-5 recommendation result (Bold font is our method).

|       | $Precision@5$ | $Recall@5$ | $F1@5$ |
|-------|---------------|------------|--------|
| KGCN  | 0.0272(-5.42%) | 0.1312(-19.69%) | 0.0451(-8.53%) |
| KGAT  | 0.0360(-4.54%) | 0.1508(-17.73%) | 0.0581(-7.23%) |
| KGIN  | 0.0690(-1.24%) | 0.2567(-7.14%)  | 0.1088(-2.16%) |
| **UGPIG** | **0.0814** | **0.3281** | **0.1304** |

**4.3.2 Impact of parameters**

In this section, we evaluate the impact of different parameters on the performance of the UGPIG method by comparing the overall performance evaluation metric, $F1@3$, for the Top-3 recommendations. Specifically, following the parameter settings mentioned in section 4.2, when studying a particular parameter, we fix the other parameters and vary the value of that parameter to observe the changes in the model's performance.

**Impact of neighbor aggregation size**：By varying the number of aggregation layers L in the User Graph, we investigated the impact of layer depth on the model's recommendation performance. As shown in Table 5, we observed that as the number of layers increased, higher-order semantic information from the User Graph was captured, and the recommendation performance gradually improved. However, a decline in performance was observed when L=5, which we attribute to the introduction of noise and perturbation caused by the repetition of path information during the process of information aggregation.

Table 5 Impact of neighbor aggregation size.

| $L$ | 2 | 3 | **4** | 5 |
|---|---|---|---|---|
| $F1@3$ | 0.1393 | 0.1375 | **0.1439** | 0.1403 |

**Impact of number of intent**：As each intent vector is independent of the others, increasing the number of intent vectors $|P|$ allows for finer granularity in capturing user preferences. Consequently, the recommendation performance improves. When the number of intent vectors becomes too large, the granularity of the intents becomes too fine, resulting in a loss of useful information. As shown in Table 6, the performance of the recommendation system starts to decline when the number of intent vectors changes from 4 to 5.

Table 6 Impact of number of intent vectors.

| $|P|$ | 3 | **4** | 5 |
|---|---|---|---|
| $F1@3$ | 0.1337 | **0.1439** | 0.1405 |

**Impact of graph pruning**：The size of the User Graph before and after pruning is shown in Table 7. As shown in Table 8, we denote the UGPIG method without graph pruning as $UGPIG_{w/o\ GP}$. It can be observed that the knowledge graph without pruning has low connectivity, and multi-hop paths among nodes can lead to a significant amount of redundant path calculations, resulting in lower performance of the recommendation system. By applying graph pruning, 38,075 geographic feature nodes were removed, resulting in a 0.854% increase in graph density. The performance of the recommendation system improved by 1.27%.

Table 7 Statistics of the UKG and UKGP.

|  | Users | Features | Relationship types | Density |
| --- | --- | --- | --- | --- |
| UG | 2596 | 39,744 | 29 | 0.009% |
| UGP | 2596 | 1669 | 29 | 0.863% |

Table 8 Impact of graph pruning.

|  | $UGPIG_{w/o\ GP}$ | UGPIG |
| --- | --- | --- |
| $F1@3$ | 0.1312 | **0.1439** |

**Impact of attention block**: As shown in Table 9, we denote the UGPIG method that does not use an attention block to fuse the features from IG and UGP as $UGPIG_{w/o\ Att}$. IG and UGP belong to different semantic spaces, and their representations of the target regions are different. If we simply add them together to fuse the representations, we will lose their respective spatial information. By using an attention block, the two semantic spaces are mapped to the same space through matrices, preserving their unique information. Compared to not using an attention block, F1@3 improved by 1.61%.

Table 9 Impact of attention block.

|  | $UGPIG_{w/o\ Att}$ | UGPIG |
| --- | --- | --- |
| $F1@3$ | 0.1277 | **0.1439** |

## 5. Discussion

In this section, we initially analyze the weight of each intent vector to relationship in UGP, which provided insights into the importance of target region features in recommending sustainable development patterns. Subsequently, we utilized Fujian Province as a case study to formulate future development directions for sustainable development patterns in the region. Moreover, we performed a detailed comparative analysis with relevant government documents.

### 5.1 Analysis of Intention Connotation

The proposed method in this paper can reveal the most relevant feature elements for each sustainable development pattern, such as culture, ecology, climate, and others. As depicted in Figure 2, each vector $p$ in the intent vector group $P$ exhibits different weights for the relationships in UGP. Table 10 below presents the top 5 relationships with the highest weights for each intent vector.

Table 10 Top-5 relationship of intent vectors.

|  | 1st relationship | 2nd relationship | 3rd relationship | 4th relationship | 5th relationship |
|---|---|---|---|---|---|
| $p1$ | Area.CultivateArea | Area.SoilErosion | Area.SecondInGDP | Area.WoodCover | Area.ThirdInGDP |
| $p2$ | Area.NatureRisk | Area.Climate | Area.WaterQuality | Area.City | Area.Province |
| $p3$ | Area.WoodCover | Area.WaterQuality | Area.WaterPer | Area.UrbanSaving | Area.NOICH |
| $p4$ | Area.GrassCoverage | Area.UrbanSaving | Area.NatureReserve | Area.Dialect | Area.BiologyHabitat |

From Table 10, we can observe that in the Top-5 relationships of $p1$, two of them are related to Richness of Resources in Table1, indicating that $p1$ has a stronger preference for Richness of Resources. For $p2$, three Geographical Location-related relationships appear in the Top-5, suggesting that $p2$ emphasizes the Geographical Location information of the target region. The Top-5 relationships of $p3$ cover

Environmental Friendliness, Geographical Location, Richness of Resources, Economic Development, and Cultural Factors, indicating a balanced intent for $p3$. In the case of $p4$, two relationships are related to Environmental Friendliness, indicating that $p4$ values the environmental information of the target region. Although each intent vector $p$ has different weights for the relationships in the UGP, by examining the frequency and order of appearance of relationship types in the Top-5 relationships, we can observe a decreasing dependency of sustainable development patterns on Environmental Friendliness, Richness of Resources, Geographical Location, Economic Development, Cultural Factors, and Social Development. Firstly, areas with higher Environmental Friendliness and Richness of Resources, as well as superior Geographical Location, often possess better natural conditions and environmental awareness, making them more conducive to the development of sustainable development patterns and ensuring the sustainability of such patterns. Therefore, areas with a better starting point are more likely to achieve sustainable development. Secondly, economically developed areas typically have more abundant human and financial resources, enabling them to invest more funds and technological support, making it easier for them to implement sustainable development patterns and achieve better results. But relatively impoverished areas may face challenges such as insufficient funding and lack of technological support, requiring more government and social support to achieve sustainable development. Furthermore, Cultural Factors and Social Development also have certain influences on the realization of sustainable development. Areas with higher cultural qualities and social awareness are more likely to implement ecological civilization construction from cultural and social perspectives. In areas with weaker cultural and social development, more efforts and time may be needed to promote sustainable development. In conclusion, the dependency of target region on Environmental Friendliness, Richness of Resources, Geographical Location, Economic Development, Cultural Factors, and Social Development gradually decreases when developing sustainable development patterns. This is the result of the comprehensive interaction among various factors, highlighting the need for tailored implementation of sustainable development patterns based on local conditions.

Below is a specific example. Figure 6 shows the Top-3 recommendations for Shangri-La City using the UGPIG method, namely $i6$ Natural Park pattern, $i15$ Characteristic Town pattern, and $i37$ Orchard Livestock pattern. Among them, the importance scores of the four intent vectors for Shangri-La City are 0.21, 0.21, 0.26, and 0.32, respectively. The score for vector $p4$ is the highest, indicating that $p4$ emphasizes the Environmental Friendliness of the target region. This aligns with the analysis results mentioned earlier, which suggest that sustainable development patterns have the highest dependency on Environmental Friendliness. The actual recommendation results in Shangri-La City support this finding. Unlike described in KGIN (X. Wang et al., 2021) where movie recommendations depend on a specific intent, the importance scores of the intent vectors in our case are very close. Although $p4$ has the highest score, it indicates that the recommendation of sustainable development patterns in geoscientific scenarios is a result of the collective influence of multiple intent vectors rather than relying on a single intent. Therefore, Shangri-La City obtains the $i6$, $i15$, and $i37$ patterns through the combined effects of the four intent vectors.

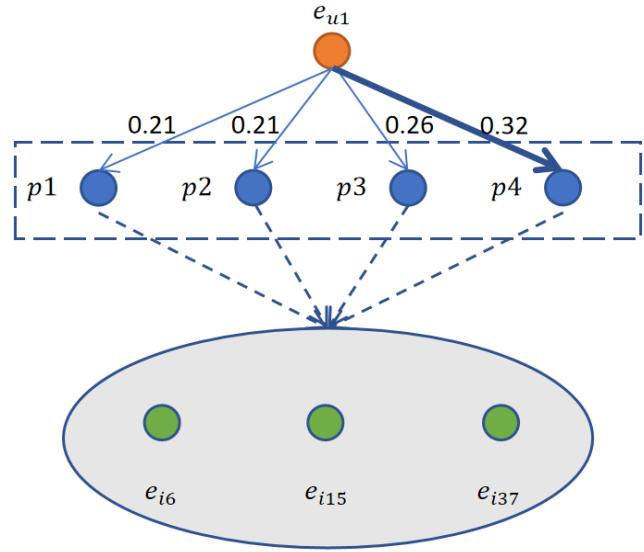

Figure 6 Top-3 Recommendation for Shangri-La City.

## 5.2 Case study

In this section, to assess the coincidence between the top-5 results recommended by the UGPIG method for counties in Fujian Province and the historical sustainable development patterns of the target regions, we can compare the second-level categories of sustainable development patterns to which they belong. Through comparison, it points out the direction for the sustainable development of Fujian Province in the future.

### 5.2.1 Sustainable development direction planning

In Section 4.1, it was mentioned that the 94 third-level sustainable development patterns can be classified into six categories of second-level sustainable development patterns. These categories are: Nature Conservation Pattern, Ecological Restoration and Governance Pattern, Ecological Agriculture Pattern, New Urbanization Pattern, Ecological Industrial Pattern, and Green Consumption Pattern. With the consideration of geographic heterogeneity, we compare the top-5 recommended results for each county in Fujian Province using the UGPIG method with the second-level sustainable development pattern categories to which the past sustainable development patterns in Fujian Province belong. If there is coincidence between the two, it indicates that the government's past decision-making direction is appropriate. However, if there is a significant deviation, adjustments need to be made for future development directions. Specifically, there are four development directions: Clearly Oriented (coincidence degree = 100%), Unhurriedly Adjustable (coincidence degree > 50%), Expectantly Transitional (coincidence degree < 50%), and Urgently Transitional (coincidence degree = 0%). The formula for calculating the coincidence degree is as follows:

$$coincidence\ degree = \frac{|E(u) \cap R(u)|}{|E(u)|} \qquad (15)$$

Consider the example of Siming District. $E(u)$ represents the set of second-level categories to which the past sustainable development patterns in Siming District belong, and $R(u)$ represents the set of second-level categories to which the top-5 recommended sustainable development patterns belong. In Siming District, the past sustainable development patterns belong to three second-level categories: Ecological Restoration and Governance Pattern, Ecological Agriculture Pattern, and Green Consumption Pattern. However, the top-5 recommended sustainable development patterns belong to the categories of Nature Conservation Pattern, Ecological Agriculture Pattern, and New Urbanization Pattern. Among these, only the Ecological Agriculture Pattern is present in both the past development list and the recommended results. Therefore, the *coincidence degree* is 1/3 = 33.33%. This indicates that Siming District's development direction is Expectantly Transitional.

The development direction planning results for the Fujian region are shown in Figure 7. Among the 84 districts and counties in Fujian Province, during the creation of the UKG, there were six regions where no historical interaction with sustainable development patterns was identified. These regions are represented as gray in Figure 7, indicating that they cannot be compared with existing representative sustainable development patterns due to the absence of historical data. Therefore, in this analysis, we consider a total sample size of 78. Overall, among these 78 samples, the proportion of districts categorized as "Clearly Oriented" and "Unhurriedly Adjustable" is 78.20%. This indicates that the majority of regions in their current development of sustainable development patterns take into account local geographical factors and follow objective development principles. However, there exists an imbalance in the development of ecological civilization at the local level. From Figure 7, it can be observed that districts classified as "Urgently Transitional" are predominantly concentrated in the northern region. Taking the example of Siming City and Ningde City, all the counties under Siming City fall into the transitional categories of "Clearly Oriented" and "Unhurriedly Adjustable," while 33.3% of the districts in Ningde City are classified as "Urgently Transitional." In Fujian Province, where sustainable development pattern development largely adheres to objective principles and exhibits an imbalance among local regions, the recommendations provided by sustainable development pattern analysis can serve as a reference for guiding sustainable development transformation in different regions.

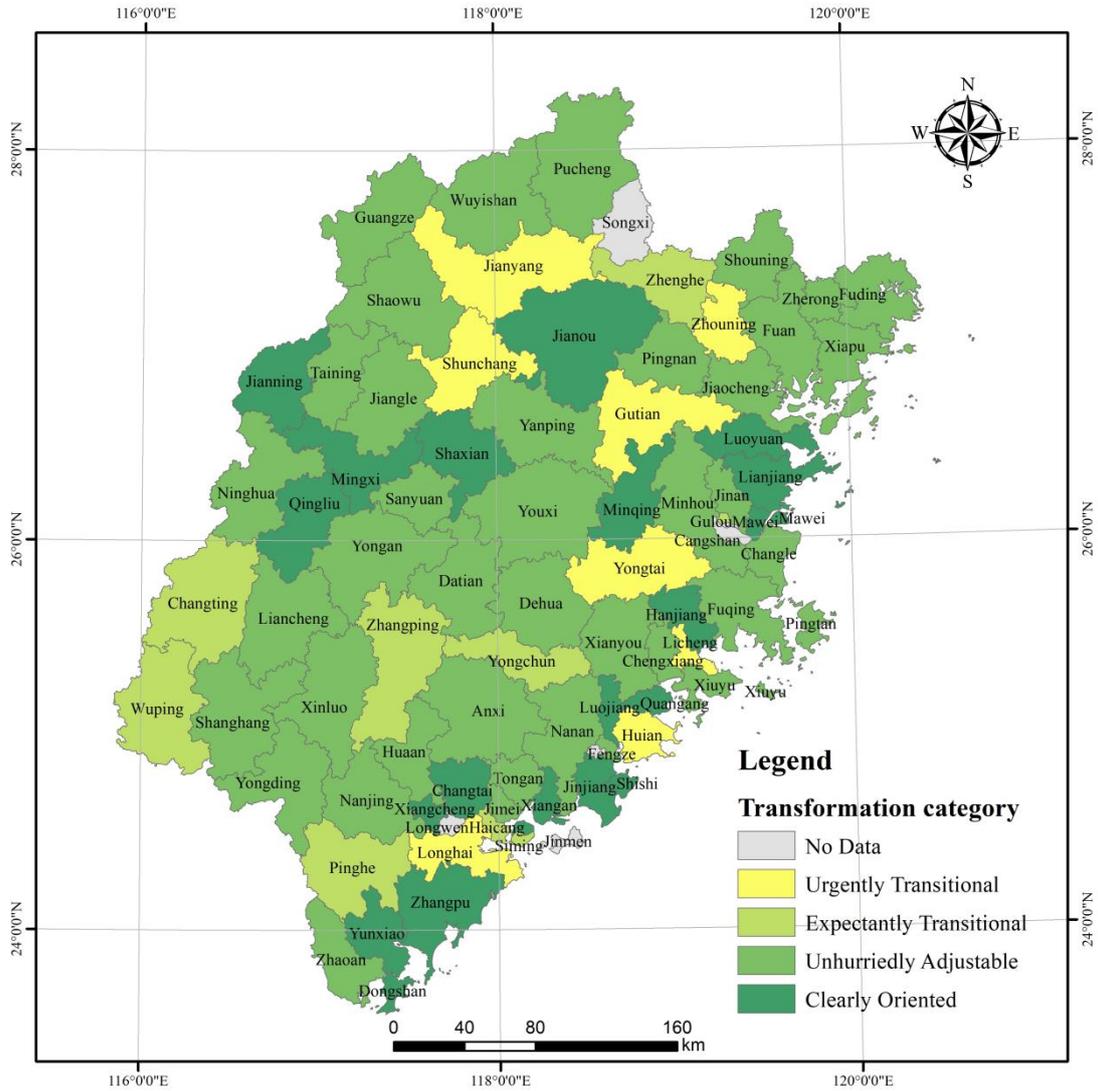

Figure 7 Sustainable development direction planning in Fujian Province.

**5.2.2 Planning analysis**

The UGPIG method effectively utilizes the recommended results for the target region and the historical sustainable development pattern categories to plan the ecological civilization development direction in Fujian Province. In this section, we will analyze the ecological development in government planning documents and calculate the accuracy of the Top-5 recommendations generated by the UGPIG method. This will provide a better understanding of the scientific validity of the development direction planning obtained through the UGPIG method.

Firstly, we define the accuracy of the Top-5 recommendations as follows:

$$Accuracy = \frac{\sum_{d \in D}|Top(d) \cap Gov(d)|}{5 * |D|} \qquad (16)$$

$D$ is the set of 78 counties in Fujian Province, $Top(d)$ represents the set of Top-5 recommendations for county $d$, and $Gov(d)$ represents the set of sustainable development patterns mentioned in the government planning documents for county d. This formula reflects the extent to which the Top-5 recommendations are reflected in the government planning documents of each region.

After calculation, the accuracy of the Top-5 recommendations in Fujian Province is 79.74%. The results provide the following insights and conclusions regarding the ecological development of Fujian Province:

•High scientific and feasibility level of government planning: The government's ecological development plans show a high degree of similarity to the recommended sustainable development patterns based on data analysis. This indicates that the government planning process has adequately considered the changing trends in ecosystems and the impact of human activities on the environment.

•High credibility of the model: By measuring the accuracy against government planning, we can validate the high credibility of our UGPIG method and further enhance the practicality of sustainable development pattern recommendations.

•Reinforcement of scientific and practical sustainable development: The similarity between government planning and the recommended sustainable development patterns suggests that the insights from government planning and the recommended sustainable development patterns can complement and promote each other in the process of ecological construction. This strengthens the scientific basis and practical effectiveness of sustainable development initiatives.

**6. Conclusion**

The proposed method in this paper, UGPIG, addresses the recommendation of sustainable development patterns. The method begins by constructing User Graph based on regional features and Intent Graph which incorporates user intent. It employs UG pruning within the knowledge graph to enhance relevance and generate UGP. Additionally, it leverages user information in both the Intent Graph and User Graph, aggregating the representations of the target region through attention mechanisms. This process generates scores for the sustainable development patterns in the target region, ultimately providing recommendations for specific areas. Experimental results demonstrate significant improvements in the performance of the sustainable development pattern recommendation model, particularly in considering geographical heterogeneity.

However, it should be noted that the current implementation of UGPIG relies solely on the relationships within the User Graph to construct the intent vector. This approach lacks the ability to provide high-quality interpretability in the form of intent vectors, which must be manually provided and cannot be automatically learned by the model. Furthermore, aggregating user information in both the Intent Graph and the User Graph introduces additional noise. Additionally, due to the lack of metadata for sustainable development patterns, enriching the representation of patterns with multimodal data such as images and text could be explored. Therefore, future research should focus on addressing these challenges, including noise reduction, automatic determination of intent vector count, and the incorporation of multimodal representations for sustainable development patterns.